\documentclass{jpp}
\usepackage{graphicx}

\usepackage[utf8]{inputenc}
\usepackage[T1]{fontenc}
\usepackage{amsmath}
\usepackage{bm}
\usepackage[dvipsnames]{xcolor}
\usepackage{physics}
\usepackage[export]{adjustbox}

\usepackage{float}
\usepackage{comment}
\usepackage{setspace}

\newcommand{\vpol}{\bm{v}_{p}}
\newcommand{\vpols}{\bm{v}_{p s}}

\newcommand{\vzero}{\bm{\bar v}}
\newcommand{\vzeros}{\bm{\bar v}_s}

\newcommand{\vperpzero}{\bm{\bar v}_{\perp}}
\newcommand{\vperpzeros}{\bm{\bar v}_{\perp s}}

\newcommand{\cycl}{\Omega_c}
\newcommand{\cycls}{\Omega_{cs}}

\newcommand{\vexb}{\bm{v}_{E}}

\newcommand{\vdiams}{\bm{v}_{* s}}
\newcommand{\vpis}{\bm{v}_{\bm{\pi}s}}
\newcommand{\vRs}{\bm{v}_{\bm{\mathcal{R}} s}}

\newcommand{\Jpol}{\bm{J}_{p}}
\newcommand{\Jpols}{\bm{J}_{p s}}

\newcommand{\Mtot}{\bm{\mathcal{\overline M}}}

\newcommand{\bvec}{\bm{b}}

\newcommand{\vpars}{\bm{v}_{\parallel s}}
\newcommand{\vperps}{\bm{v}_{\perp s}}

\newcommand{\Jperpzero}{\bm{\bar J}_\perp}

\newcommand{\Jpar}{\bm{J}_\parallel}

\shorttitle{Conservative formulation of the drift-reduced plasma model}
\shortauthor{B. De Lucca et al.}

\title{Conservative formulation of the drift-reduced fluid plasma model}

\author{B. De Lucca\aff{1}
  \corresp{\email{brenno.delucca@epfl.ch}},
  P. Ricci\aff{1},
  M. Bassanini\aff{1,2},
  S. García Herreros\aff{1},\\
  Z. Tecchiolli\aff{1}
}

\affiliation{\aff{1}École Polytechnique Fédérale de Lausanne (EPFL), Swiss Plasma Center (SPC), CH-1015 Lausanne, Switzerland 
\aff{2}École Polytechnique Fédérale de Lausanne (EPFL), Institute of Mathematics, CH-1015 Lausanne, Switzerland 
}

\begin{document}

\maketitle

\begin{abstract}
A conservative formulation of the drift-reduced fluid plasma model is constructed by analytically inverting the implicit relation defining the polarisation velocity as a function of the time-derivative of the electric field. The obtained model satisfies exact conservation laws for energy, mass, charge and momentum, in arbitrary magnetic geometry, also when electromagnetic fluctuations are included.
\end{abstract}

\section{Introduction} \label{sec:Intro} 
 In highly-collisional regimes, magnetised plasma turbulence is often studied with drift-reduced fluid models (see, e.g., \cite{drake_antonsen_1984, zeilerdrakerogers_DR_1997, scott_2003, simakov_catto_2003}). These models are commonly employed to investigate the boundary region of fusion devices~\citep{Bout_Dudson_2009, Zhu_GDB, Stegmeir_2019, Giacomin_2022, Dull_2024} and also find applications in basic plasma physics experiments~\citep{Riva_2016}.
 
 The drift-reduced formulation of the fluid plasma is derived by perturbatively expanding the fluid moment equations in powers of the drift-expansion parameter $\epsilon \sim d_t/\cycl \ll 1$, which represents the ratio between the dynamical timescale, encoded in the material derivative $d_t$, to the cyclotron frequency $\cycl = q B/m$, expressed in terms of the charge $q$, the mass $m$, and the magnetic field strength $B = |\bm{B}|$. In the limit $\epsilon \ll 1$, the projection of the fluid velocity in the plane orthogonal to the magnetic field, $\bm{v}_\perp$, decomposes in a set of drift contributions. Among these, the polarisation drift $\vpol \sim (B \cycl)^{-1} d_t \bm{E}$, is related to the time and spatial evolution of the electric field $\bm{E}$. This allows determining the electric field evolution through a vorticity equation. Indeed, by involving $\vpol$, the quasi-neutrality constraint $\nabla \cdot \bm{J} = 0$ results in an equation for the electrostatic potential. However, the drift-reduced models as currently expressed in the literature and implemented in the high-fidelity fluid codes for tokamak and stellarator turbulence simulations~\citep{Bout_Dudson_2009, Stegmeir_2019, Giacomin_2022, Coelho_2024, Dull_2024}, are known to lack exact conservative properties, with spurious source terms appearing at $O(\epsilon)$~\citep{reiserAIP_2012, halpern_energy2023}.
 
To derive a conservative drift-reduced fluid model, it is necessary to invert an implicit equation to relate the polarisation velocity to the time-derivative of the electric field, where formally small terms need to be kept to ensure conservation~\citep{zeilerdrakerogers_DR_1997, reiserAIP_2012, poulsen_ramussen_multispecies, halpern_energy2023}. This was done in the geometry of a linear device for a cold-ion plasma in the electrostatic limit by~\citet{reiserAIP_2012}, by exploiting the cylindrical symmetry of the background magnetic field to derive an explicit time-evolution equation for the vorticity. In this work, we derive a closed-form, non-perturbative expression for the polarisation velocity 
$\vpol = \vpol(\partial_t \bm{E})$, in an arbitrary magnetic
geometry and without imposing the electrostatic limit, and use it to build an exactly energy and momentum conserving drift-reduced fluid model. The construction holds for an arbitrary number of species and does not depend on the choice of fluid closure. 

As drift-reduced fluid codes are becoming tools of reference in building predictive capabilities for the operation and design of fusion devices~\citep{GiacominPRL_2022, OliveiraTCVX21_2022, Bufferand_2024, Zholobenko_2024}, the formulation of the drift-reduced model in a conservative form is relevant for several reasons. First, the existence of exact invariants is important when devising numerical schemes to increase code performance~\citep{LeVeque1992}. Second, it proves that the construction of a drift-reduced model obeying the same conservation properties of the non drift-reduced system is possible in general, implying that this approximation scheme for fluid magnetised plasmas is well-posed. 

This paper is organised as follows. Following the Introduction, in Sec~\ref{sec:starting_fluid_model} we express the fluid equations for a multispecies quasi-neutral magnetised plasma and explicit the corresponding energy and momentum conservation laws. In Sec~\ref{sec:canonical_drift_reduction}, we state the drift-approximation ordering and explain why a perturbative expansion of the polarisation velocity as a function of leading-order quantities breaks the model's conservation properties. In Sec~\ref{sec:inversion_vpol}, the implicit and non-perturbative expression for the polarisation drift is inverted analytically, leading to an exactly conservative model. In Sec~\ref{sec:conservation_proof}, the $O(\epsilon)$ drift-reduced model is given and shown to conserve the leading-order components of energy and momentum exactly (i.e., to all orders in $\epsilon$) in Sec~\ref{sec:conservation_properties}. Finally, the conclusions are drawn in Sec~\ref{sec:Conclusion}.

 


\section{Fluid equations of a multispecies plasma}\label{sec:starting_fluid_model}

We consider a magnetised plasma containing an arbitrary number of distinct species, indexed by the label $s$. 
 The fluid equations for a given species $s$, of mass $m_s$ and electric charge $q_s$, are obtained by taking velocity moments of the Boltzmann equation~\citep{Braginskii_1965}. The first three moment equations, evolving the density $n_s$, the fluid momentum density $\bm{\mathcal{M}}_s = m_s n_s \bm{V}_s$, with $\bm{V}_s$ the mean velocity of species $s$, and the scalar pressure $p_s = n_s T_s$, with $T_s$ the temperature of species $s$, are given by 
\begin{align}
    &\frac{\partial n_s}{\partial t} + \nabla \cdot (n_s \bm{V}_s) = \mathcal{S} _s, \label{eqn:continuity} \\
    &\frac{\partial \bm{\mathcal{M}}_s}{\partial t} + \nabla \cdot (\bm{V}_s \bm{\mathcal{M}}_s) + \nabla \cdot \bm{P}_s - q_s n_s\left(\bm{E} + \bm{V}_s \times \bm{B}\right) = \bm{\mathcal{R}}_s ,\label{eqn:momentum}\\
    &\frac{\partial}{\partial t}\left(\frac{3}{2} p_s \right) + \nabla \cdot \left(\frac{3}{2}p_s \bm{V}_s + \bm{q}_s \right) + \bm{P}_s : \nabla \bm{V}_s= \mathcal{Q}_s + r_s \mathcal{K}_s , \label{eqn:pressure}
\end{align}
where $\bm{P}_s = p_s \mathbb{I} + \bm{\pi}_s$ is the pressure tensor of the species, $\bm{\pi}_s$ is the stress-tensor, $\mathbb{I}$ is the identity matrix, $\bm{q}_s$ is the heat-flux density and $\mathcal{S}_s, \bm{\mathcal{R}}_s, \mathcal{Q}_s$ are source terms, arising from external drives and interactions between the species, e.g. via Coulomb collisions. In addition, we define $r_s \equiv \mathcal{S}_s/n_s$ as the rate at which particles of species $s$ enter or leave the system and $\mathcal{K}_s \equiv m_s n_s V_s^2/2$ is the fluid kinetic energy density of species $s$. In Eqs~\ref{eqn:momentum} and \ref{eqn:pressure}, the notation $\bm{P} : \nabla \bm{V} \equiv P^{ij} \nabla_j V_{i}$ is introduced to represent the Frobenius product of the two tensors and the divergence of a tensor $\nabla \cdot \bm{P}_s \equiv \nabla_i P^{ij}$ is defined as contracting over neighbouring indices. We express the application of a tensor on a vector with a similar abuse of notation and write $\bm{P} \cdot \bm{V} \equiv P_{ij} V^j$, with the tensorial property of the resulting quantity discernible from the context. For the sake of generality, we do not focus on the closure of the moment hierarchy in this work. We assume that an adequate closure is encoded in the heat-flux densities and stress-tensors, as well as the contributions to the source terms arising from interparticle interactions. We note that the particle sources $\mathcal{S}_s$ cannot all be chosen independently, as charge-neutrality imposes the constraint $\sum_s q_s \mathcal{S}_s = 0$.   

Since the turbulent phenomena we wish to model tend to evolve on scales much larger than the Debye length, $\lambda_D$, and timescales much slower than the plasma frequency, $\omega_p$, we assume
\begin{align}
    \lambda_D \nabla \sim \frac{1}{\omega_p}\frac{\partial}{\partial t} \equiv \epsilon_D \to 0 .
\end{align}
In this limit, the plasma is instantaneously locally quasi-neutral and the electric field $\bm{E}$ is not determined by Gauss' law but rather by the quasi-neutrality condition $\nabla \cdot \bm{J} = 0$.
Thus, Maxwell's equations in the limit $\epsilon_D \to 0$ reduce to
\begin{align}
    \nabla \times \bm{B} &= \mu_0 \bm{J} = \mu_0 \sum_s q_s n_s \bm{V}_s ,\label{eqn:ampere} \\
    \nabla \times \bm{E} &= - \frac{\partial \bm{B}}{\partial t} ,\label{eqn:faraday}\\
    \nabla \cdot \bm{B} &= 0 ,\label{eqn:B_solenoidal} \\
    \nabla \cdot \bm{J} &= 0 .\label{eqn:quasi_neutral}
\end{align}
Maxwell's equations, coupled to the fluid equations, Eqs~\ref{eqn:continuity}-\ref{eqn:pressure}, yield a closed model which conserves energy, momentum, mass and charge. 

The energy density $\mathcal{H}$ can be expressed as
\begin{align}\label{eqn:full_energy_decomp}
    \mathcal{H} = \mathcal{H}_B + \sum_s \mathcal{H}_s = \frac{B^2}{2 \mu_0} +  \sum_s \left(\mathcal{K}_s + \frac{3}{2}p_s\right) ,
\end{align}
where $\mathcal{H}_s = \mathcal{K}_s + \mathcal{U}_s$ is the energy of each species, composed of the kinetic energy, $\mathcal{K}_s$, and the internal energy, $\mathcal{U}_s \equiv 3 p_s/2 $, while $\mathcal{H}_B \equiv B^2/2 \mu_0$ is the field energy, which contains only the magnetic energy density contribution in the quasi-neutral limit.
To demonstrate the conservation of $\mathcal{H}$, we take the scalar product of the momentum equation, Eq~\ref{eqn:momentum}, with $\bm{V}_s$ and use the continuity equation, Eq~\ref{eqn:continuity}. This leads to a transport equation for the fluid kinetic energy density $\mathcal{K}_s$
\begin{align}\label{eqn:kinetic_energy}
    \frac{\partial \mathcal{K}_s}{\partial t} + \nabla \cdot \left(\mathcal{K}_s\bm{V}_s \right)  &= - \bm{V}_s \cdot \nabla \cdot \bm{P}_s +   q_s n_s \bm{V}_s \cdot \bm{E}  +  \bm{V}_s \cdot \bm{\mathcal{R}}_s - \mathcal{K}_s\frac{\mathcal{S}_s}{n_s} .
\end{align}
Furthermore, we note that the electromagnetic field energy evolves according to Poynting's theorem and, in a quasi-neutral magnetised plasma, where $\omega/\omega_p \to 0$, the term due to the electric field energy, $\partial_t (\epsilon_0 E^2/2)$, is negligible. Indeed, the displacement current scales as $\partial_t \bm{E}/c^2 \sim (\omega \Omega_{ce}/\omega_{pe}^2) \mu_0 \bm{J}$, where the electron timescales are the dominant ones, so in the neutral limit $\omega/\omega_{pe} \to 0$, we have that $\partial_t \bm{E}/c^2 \to 0$ so long as $\Omega_{ce}/\omega_{pe}$ does not diverge  (which almost always holds in magnetised plasmas). In this limit, Poynting's theorem is given by
\begin{align}\label{eqn:poynting_qn}
    \frac{\partial \mathcal{H}_B}{\partial t} + \nabla \cdot \bm{S} + \bm{E} \cdot \bm{J} &= 0,
\end{align}
with $\bm{S} \equiv \bm{E} \times \bm{B}/ \mu_0$ the Poynting vector, as it can be observed by taking the scalar product of Faraday's law with $\bm{B}$ and using the fact that $\bm{B} \cdot (\nabla \times \bm{E}) = \mu_0 \nabla \cdot \bm{S} + \bm{E} \cdot (\nabla \times \bm{B})$, leading to
\begin{align}\label{eqn:HB_part}
    \frac{\partial }{\partial t}\left(\frac{B^2}{2 \mu_0}\right) + \nabla \cdot \bm{S} + \frac{1}{\mu_0}\bm{E} \cdot (\nabla \times \bm{B}) = 0,
\end{align}
and then imposing $\nabla \times \bm{B} = \mu_0 \bm{J}$.
Thus, summing the transport equations for the internal energy, Eq~\ref{eqn:pressure}, the fluid kinetic energy, Eq~\ref{eqn:kinetic_energy}, the magnetic field energy, Eq~\ref{eqn:poynting_qn}, results in an equation for the total energy density
\begin{align}\label{eqn:energy_full_system}
    \frac{\partial \mathcal{H}}{\partial t} &+ \nabla \cdot \left[\bm{S} + \sum_s (\mathcal{H}_s \bm{V}_s + \bm{P}_s \cdot \bm{V}_s + \bm{q}_s) \right] = \sum_s \left(\mathcal{Q}_s + \bm{V}_s \cdot \bm{\mathcal{R}}_s \right) ,
\end{align}
where a further summation over all the species $s$ has been performed and the symmetry property of the pressure tensor $\bm{P}_s = \bm{P}^T_s$ is used to write $\bm{P}_s \cdot \nabla \cdot \bm{P}_s + \bm{P}_s : \nabla \bm{V}_s = \nabla \cdot (\bm{P}_s \cdot \bm{V}_s)$ in Eq~\ref{eqn:energy_full_system}. 
 The energy content of the system changes due to heat sources, $\mathcal{Q}_s$, and forces, $\bm{V}_s \cdot \bm{\mathcal{R}}_s$. Defining the total power source $\mathcal{S}_\mathcal{H} = \sum_s(\mathcal{Q}_s + \bm{V}_s \cdot \bm{\mathcal{R}}_s)$ and the energy flux $\bm{\Gamma}_\mathcal{H} \equiv \bm{S} + \sum_s (\mathcal{H}_s \bm{V}_s + \bm{P}_s \cdot \bm{V}_s + \bm{q}_s)$, Eq~\ref{eqn:energy_full_system} can be written as a conservation law for $\mathcal{H}$,
\begin{align}
    \frac{\partial \mathcal{H}}{\partial t} + \nabla \cdot \bm{\Gamma}_{\mathcal{H}} = \mathcal{S}_\mathcal{H}.
\end{align}
The source term $\mathcal{S}_\mathcal{H}$ vanishes if the system is not externally driven and particle interactions are elastic, as is the case for the Coulomb interaction~\citep{Braginskii_1965}.
A transport equation can be similarly derived for the momentum density $\bm{\mathcal{M}} \equiv \sum_s \bm{\mathcal{M}}_s$
\begin{align}\label{eqn:total_momentum_density}
    \frac{\partial \bm{\mathcal{ M}}}{\partial t} + \nabla \cdot \sum_s (\bm{V}_s \bm{\mathcal{ M}}_s + \bm{P}_s)  &=  \bm{J} \times \bm{B} + \sum_s \bm{\mathcal{R}}_s ,
\end{align}
where, in the presence of elastic interactions and in the absence of external momentum input, the term $\sum_s \bm{\mathcal{R}}_s$ vanishes. Total mass is conserved by virtue of the continuity equation and finally, charge is conserved by the requirement that $\nabla \cdot \bm{J} = 0$. 

\section{Drift-approximation}\label{sec:canonical_drift_reduction}

We focus on low-frequency dynamics, therefore assuming that the physics of interest evolves on a scale of the order of the drift-scale, $\partial_t \sim \bm{V}_s \cdot \nabla \sim \omega_{* s} \equiv \bm{k}_\perp \cdot \vdiams$, where $\vdiams = \bm{B} \times \nabla p_s/(q_s n_s B^2)$ is the diamagnetic drift-velocity of species $s$ and $\bm{k}_\perp$ is the wavenumber in the plane orthogonal to $\bm{B}$~\citep{zeilerdrakerogers_DR_1997}.  The drift-scale is assumed much slower than the cyclotron frequency $\omega_{* s}  \ll \cycls$, thus $\partial_t/\cycls \sim  \bm{ V}_s \cdot \nabla/\cycls \equiv \epsilon$, with $\epsilon \ll 1$ a small parameter. This implies that the fluid velocities and length scales are ordered as $V_{\perp s} k_\perp \sim V_{\parallel s} k_\parallel \sim \epsilon \cycls$, where $k_\perp = |\bm{k}_\perp|$ and $k_\parallel = \bm{k} \cdot \bvec$ denote the perpendicular and parallel wavenumbers, and $V_{\perp s}$ and $V_{\parallel s}$ represent the corresponding perpendicular and parallel components of the fluid velocity.
Our ordering choice is consistent with both long-wavelength/large-flow drift-reduced assumptions, $k_\perp \rho_{Ls} \sim \sqrt{\epsilon}, V_{\perp s} \sim \sqrt{\epsilon} v_{Ts}$, and short-wavelength/small-flow gyrokinetics, $k_\perp \rho_{Ls} \sim 1, V_{\perp s} \sim \epsilon v_{Ts}$, where $v_{Ts} = \sqrt{T_s/m_s}$ is the thermal velocity of species $s$, and $\rho_{Ls} = v_{Ts}/\cycls$ its Larmor radius~\citep{Brizard_Hahm_2007}. Indeed, here we adopt a more general ordering than the one typically assumed in drift-reduced models, in which only large-scale dynamics are retained $k_\perp \rho_{Ls} \sim \sqrt{\epsilon}$~\citep{drake_antonsen_1984, zeilerdrakerogers_DR_1997, simakov_catto_2003}, as the large-scale assumption is not required to derive the results presented in this work. 

In the $\epsilon = 0$ limit, key physical processes that give rise to electrostatic $E \times B$ turbulence are absent~\citep{zeilerdrakerogers_DR_1997}. In order to retain them in our model, we expand the fluid equations, Eqs~\ref{eqn:continuity}-\ref{eqn:pressure}, in powers of $\epsilon$~\citep{drake_antonsen_1984}. Decomposing the fluid velocity $\bm{V}_s$ of species $s$ in terms of its parallel and perpendicular components with respect to the magnetic field $\bm{B}$  
\begin{align}
    \bm{V}_s = \vpars + \vperps ,
\end{align}
and taking the cross product of the momentum equation, Eq~\ref{eqn:momentum}, with $\bm{B}$, we find that $\vperps$ decomposes into the sum of drift contributions~\citep{drake_antonsen_1984}
\begin{align}
    \vperps = \vexb + \vdiams + \vpis + \vRs + \vpols,
\end{align}
that is an $E\times B$ drift term, $\vexb = \bm{E} \times \bm{B}/B^2$, a diamagnetic drift related to isotropic pressure, $\vdiams$, a diamagnetic drift due to pressure anisotropies, $\vpis = \bm{B} \times \nabla \cdot \bm{\pi}_s/(q_s n_s B^2)$, a drift due to momentum source/sinks, $\vRs = \bm{\mathcal{R}}_s \times \bm{B}/(q_s n_s B^2)$, and a drift related to the rate of change of the bulk-flow, the polarisation velocity $\vpols$, here defined as
\begin{equation}\label{eqn:vpol_def}
    \vpols \equiv \frac{\bvec}{n_s \cycls} \times \left[\frac{\partial (n_s\bm{V}_s)}{\partial t} + \nabla \cdot (n_s \bm{V}_s \bm{V}_s) \right] ,  
\end{equation}
where $\bvec = \bm{B}/B$ is the unit vector in the direction of $\bm{B}$. Using the continuity equation, Eq~\ref{eqn:continuity}, the polarisation velocity can also be expressed as
\begin{align}\label{eqn:vpol_def_2}
    \vpols = \frac{\bvec}{ \cycls} \times \left(\frac{d_s \bm{V}_s}{dt} + r_s \bm{V}_s\right),
\end{align}
where $d_s/dt \equiv \partial_t + \bm{V}_s \cdot \nabla$ is the material derivative of species $s$.  The polarisation drift results from the particle finite inertia and is subleading in the imposed drift-ordering, that is $\vpols \sim (d_t/\cycls) \bm{V}_s = O(\epsilon \bm{V}_s)$, if the rate of particle injection is of the order of the drift-frequency $r_s/\cycls \sim \epsilon$, which we assume to hold. 
We can therefore write the total fluid velocity
 \begin{align}
    \bm{V}_s = \vzeros + \vpols = (\vpars + \vperpzeros) + \vpols,
\end{align}
 as the sum of a leading-order flow, $\vzeros$, and the subleading correction, $\vpols$, with $\vpars, \vzeros$ and $ \vperpzeros$ scaling as $O(\bm{V}_s)$. Explicitly, the leading-order perpendicular flow is given by
\begin{align}\label{eqn:leading_order_vperp_decomp}
    \vperpzeros = \vexb + \vdiams + \vpis + \vRs \equiv \vexb + \bm{w}_{\perp s},
\end{align}
where $\bm{w}_{\perp s}$ is the leading order drift-velocity of species $s$ in the comoving $E \times B$ frame.
 Separating the fluid velocity $\bm{V}_s$ appearing in Eq~\ref{eqn:vpol_def_2} into its leading and subleading components, one obtains
\begin{equation}\label{eqn:vpol_non_perturbative_definition}
    \vpols = \frac{\bvec}{\cycls} \times \left(\frac{\partial \vzeros}{\partial t} + \vzeros \cdot \nabla \vzeros + r_s \vzeros + \vpols \cdot \nabla \vzeros + \frac{\partial \vpols}{\partial t} + \vzeros \cdot \nabla \vpols + r_s \vpols + \vpols \cdot \nabla \vpols \right).  
\end{equation}
Eq~\ref{eqn:vpol_non_perturbative_definition} yields an implicit equation for $\vpols$, which is
\begin{equation}\label{eqn:ordinary_perturb_series_vpol}
     \vpols = F(\vzeros) + G(\vzeros, \vpols) + K(\vpols),
\end{equation}
where
\begin{align}
    F(\vzeros) &=  \frac{\bvec}{\cycls} \times \left(\frac{\partial \vzeros}{\partial t} + \vzeros \cdot \nabla \vzeros + r_s \vzeros \right) = O(\epsilon \vzeros), \\
    G(\vzeros, \vpols) &= \frac{\bvec}{\cycls} \times \left(\vpols \cdot \nabla \vzeros + \frac{\partial \vpols}{\partial t} + \vzeros \cdot \nabla \vpols + r_s \vpols \right) = O(\epsilon \vpols \vzeros), \\
    K(\vpols) &= \frac{\bvec}{\cycls} \times (\vpols \cdot \nabla \vpols) = O(\epsilon \vpols \vpols).
\end{align}
Traditionally, Eq~\ref{eqn:ordinary_perturb_series_vpol} is treated with ordinary perturbation theory $\vpols = \vpols^{(1)} + \vpols^{(2)} + \dots$, with $\vpols^{(N)} \equiv O(\epsilon^N \vzeros)$. For example, solving Eq~\ref{eqn:ordinary_perturb_series_vpol} order by order, up to $O(\epsilon^3 \vzeros)$, yields
\begin{align}
    \vpols^{(1)} &= F(\vzeros) = \frac{\bvec}{\cycls} \times \left(\frac{\partial \vzeros}{\partial t} + \vzeros \cdot \nabla \vzeros +  r_s\vzeros\right),\label{eqn:order_by_order_result}  \\
    \vpols^{(2)} &= G(\vzeros, \vpols^{(1)}) = \frac{\bvec}{\cycls} \times \left(\vpols^{(1)} \cdot \nabla \vzeros + \frac{\partial \vpols^{(1)}}{\partial t} + \vzeros \cdot \nabla \vpols^{(1)} + r_s \vpols^{(1)} \right) , \label{eqn:second_order_result}\\
    \vpols^{(3)} &= G(\vzeros, \vpols^{(2)}) + K(\vpols^{(1)}) =  G(\vzeros, \vpols^{(2)}) + \frac{\bvec}{\cycls} \times(\vpols^{(1)} \cdot \nabla \vpols^{(1)}). 
\end{align}
The resulting drift-reduced model is presented in detail in~\cite{gath_wiesenberger_AIP_2019}. Truncating the expansion Eq~\ref{eqn:order_by_order_result} to $O(\epsilon \vzeros)$, yields the models currently implemented in the drift-reduced fluid codes~\citep{Giacomin_2022, Stegmeir_2019, Dull_2024, Bout_Dudson_2015}. 

It is a known result that the models, based on the perturbative expansion in Eq~\ref{eqn:order_by_order_result}, are not conservative for both energy and momentum, with spurious source terms of $O(\epsilon)$ appearing in their time evolution~\citep{reiserAIP_2012, halpern_energy2023}. The reason is that the expression for $\vpols$ in Eq \ref{eqn:order_by_order_result} encodes the transport of the leading-order component of  perpendicular momentum, which lacks advection by the polarisation drift (a subleading term, see Eq \ref{eqn:second_order_result}). On the other hand, the polarisation advection term is retained in other parts of the model, for instance in the continuity equation, which is expanded to $O(\epsilon)$. The resulting inconsistency between the transport of momentum and density breaks the energy and momentum conservation properties of the model~\citep{reiserAIP_2012}. We further note that, given the recursive nature of the perturbative expansion method, exact conservation is not retrieved by expanding the model to higher order in $\epsilon$. Indeed, for the drift-reduced model expanded to $O(\epsilon^N)$ with $N \geq 1$, energy and momentum conservation is broken by terms of the same order, namely $O(\epsilon^{N})$.  

The solution to the conservation problem of the drift-approximation has been known since the original derivation of the model~\citep{drake_antonsen_1984, zeilerdrakerogers_DR_1997, scott_2003} and consists in retaining the polarisation advection term $\bvec \times (\vpols^{(1)} \cdot \nabla \vzeros)/\cycls$ in the perpendicular momentum transport equation, Eq~\ref{eqn:order_by_order_result}. More precisely, the polarisation drift $\vpols^{(1)}$ should be defined implicitly as
\begin{align}\label{eqn:vpol_closure}
    \vpols^{(1)} = \frac{\bvec}{\cycls} \times \left(\frac{\partial \vzeros}{\partial t} + \vzeros \cdot \nabla \vzeros  + \vpols^{(1)} \cdot \nabla \vzeros +  r_s \vzeros \right) \equiv \frac{\bvec}{\cycls} \times \left(\frac{d_s^{(1)} \vzeros}{dt} + r_s \vzeros \right),
\end{align}
instead of the perturbative definition in Eq~\ref{eqn:order_by_order_result}. The expression for $\vpols$ in Eq~\ref{eqn:vpol_closure} results in conservation laws for the leading-order contributions to energy and momentum, valid to all orders in $\epsilon$. In the following section, we first invert Eq~\ref{eqn:vpol_closure} to express $\vpols^{(1)}$ as a function of only the leading-order flow $\vzeros$ (Sec~\ref{sec:inversion_vpol}). This was the main obstacle to overcome in obtaining an explicit and exactly conservative drift-reduced model. We then derive in Sec~\ref{sec:conservation_proof} a conservative drift-reduced model. Finally, mirroring previous work on the topic~\citep{drake_antonsen_1984, zeilerdrakerogers_DR_1997, reiserAIP_2012}, we demonstrate the conservation properties of the obtained drift-reduced system in Sec~\ref{sec:conservation_properties}. Since we consider the expansion of the model only to first order in $\epsilon$, we suppress henceforth the superscript in $\vpols^{(1)}$ to lighten the notation in the rest of the paper.

\section{Solution of the implicit equation for $\vpols$}\label{sec:inversion_vpol}

 To the best of our knowledge, the expression of $\vpols$, given by the solution of Eq~\ref{eqn:vpol_closure}, has never been used in simulation codes, except for the case of an electrostatic cold-ion plasma in linear geometry~\citep{reiserAIP_2012}. In that case, the symmetry properties of the magnetic field were exploited to obtain a vorticity equation and close the system equations. In general, however, the lack of an explicit form for the polarisation velocity as a function of the electric potential's rate of change precludes the direct use of Eq~\ref{eqn:vpol_closure}. In fact, in the absence of an explicit relation $\vpols = \vpols(\partial_t \phi)$, imposing quasi-neutrality through $\nabla \cdot \bm{J} = 0$ does not yield a scalar equation for $\phi$. As a result, closing the drift-reduced fluid system requires explicit time evolution of the polarisation drift (on the cyclotron timescale), thereby defeating the purpose of the drift-reduced approximation~\citep{halpern_energy2023}. In~\citet{scott_2003}, the closure issue was avoided by assuming that the polarisation drift can be expressed in terms of the gradient of a scalar, but this relation does not hold in general. 
 
 In this section we seek to invert the expression for $\vpols$, given by Eq~\ref{eqn:vpol_closure}, as a function of the leading order velocity $\vzeros$. Reordering terms in Eq~\ref{eqn:vpol_closure}, we note that $\vpols$ satisfies
\begin{align}\label{eqn:Gpol_start}
    \vpols - \frac{\bvec}{\cycls} \times (\vpols \cdot \nabla) \vzeros = \frac{\bvec}{\cycls}\times \left(\frac{\partial \vzeros}{\partial t} + \vzeros \cdot \nabla \vzeros + r_s \vzeros  \right) .
\end{align}
We now show that the solution of Eq~\ref{eqn:Gpol_start} yields
\begin{equation}\label{eqn:vpol_solution_full_explicit}
    \vpols = \frac{1 + (\bvec \times \nabla \vzeros)/\cycls }{1 + \bvec \cdot (\nabla \times \vzeros)/\cycls + \text{det}(\nabla \vzeros)_\perp/\cycls^2} \cdot \left[ \frac{\bvec}{\cycls}\times \left(\frac{\partial \vzeros}{\partial t} + \vzeros \cdot \nabla \vzeros + r_s \vzeros  \right) \right].
\end{equation}
For convenience, we define the inertial drift-velocity term, $\bm{U}_s$, to represent the right-hand-side of Eq~\ref{eqn:Gpol_start}, that is
\begin{align}\label{eqn:R_def}
    \bm{U}_s \equiv \frac{\bvec}{\cycls}\times \left(\frac{\partial \vzeros}{\partial t} + \vzeros \cdot \nabla \vzeros + r_s \vzeros  \right), 
\end{align}
which contains only leading-order quantities related to the velocity $\vzeros$. We note that $\bm{U}_s \cdot \bvec =0$. 
Eq~\ref{eqn:Gpol_start} can be written in component form as
\begin{align}\label{eqn:Gpol_matrix_start}
     \left(\delta^i_l - \frac{1}{\cycls}\epsilon^{i j}_{\,\, \,\,k} b_j \nabla_l  \bar v_{s}^k \right)v_{ps}^l \equiv  (Q^{-1}_s)^i_{\, \, l}v_{ps}^l = U^i_s,
\end{align}
where $v_{ps}^l$ are the contravariant components of $\vpols$ in an arbitrary coordinate system, $\bm{Q}^{-1}_s \equiv (Q^{-1}_s)^i_{\, \, l}$ is the linear operator we seek to invert, $\epsilon^{ijk}$ is the Levi-Civita tensor in $\mathbb{R}^3$ and $\delta_{ij} = \delta^{ij} = \delta^i_j$ is the Kronecker delta, used to raise/lower indices in $\mathbb{R}^3$. Einstein's convention of summation over repeated indices is used throughout the paper. We simplify the calculation by considering a basis composed by $\bm{\hat w}$, an arbitrary unit vector such that  $\bm{\hat w} \cdot \bvec = 0$, $ \bvec \times \bm{\hat w}$, and $\bvec$. To lighten the notation we define $\bvec \times \bm{\hat w} \equiv \bm{\hat w}^*$, and note that $\bm{\hat w} \cdot \bm{\hat w}^* = 0$, $ \hat w^{* 2} = \hat w^2 = 1$ and $(\bm{\hat w}^*)^* = \bvec \times (\bvec \times \bm{\hat w}) = - \bm{\hat w}$. Since the polarisation velocity lies in the plane orthogonal to $\bvec$, we have 
\begin{align}\label{eqn:vpol_proj}
    \vpols = \alpha_s \bm{\hat w}  + \beta_s \bm{\hat w}^*,
\end{align}
where $\alpha_s$ and $ \beta_s$ are two scalar functions of $\bm{U}_s$ to be determined. Substituting this expression for $\vpol$ in Eq~\ref{eqn:Gpol_start}, we have
\begin{align}\label{eqn:basis_expansion_Us}
     \alpha_s \bm{\hat w} + \beta_s \bm{\hat w}^*  - \alpha_s  \frac{\bvec }{\cycls}\times (\bm{\hat w}\cdot \nabla \vzeros) - \beta_s \frac{\bvec}{\cycls}\times (\bm{\hat w}^* \cdot \nabla \vzeros) = \bm{U}_s.
\end{align}
Projecting Eq~\ref{eqn:basis_expansion_Us} onto the basis elements $\bm{\hat w}$ and $ \bm{\hat w}^*$, we obtain
\begin{align}
    \bm{\hat w} \cdot \bm{U}_s &= \alpha_s    \left(1-  \bm{\hat w} \cdot \left[\frac{\bvec}{\cycls}  \times (\bm{\hat w} \cdot \nabla \vzero)\right]\right) - \beta_s  \bm{\hat w} \cdot \left[\frac{\bvec}{\cycls} \times \left(\bm{\hat w}^* \cdot \nabla \vzeros\right)\right] ,\label{eqn:wR_eq1}\\
    \bm{\hat w}^* \cdot \bm{U}_s &= \beta_s\left(1 -\bm{\hat w}^* \cdot \left[\frac{\bvec}{\cycls} \times (\bm{\hat w}^* \cdot \nabla \vzeros)\right]\right)   - \alpha_s \bm{\hat w}^* \cdot \left[\frac{\bvec}{\cycls}  \times (\bm{\hat w} \cdot \nabla \vzeros)\right] .  \label{eqn:wR_eq2} 
\end{align}
Eqs~\ref{eqn:wR_eq1} and~\ref{eqn:wR_eq2} constitute a linear system for the unknowns $\alpha_s$ and $\beta_s$. Solving for $\alpha_s$ and $\beta_s$ yields
\begin{align}
    \alpha_s &= \frac{(1-\tilde \delta_s) \bm{\hat w} \cdot \bm{U}_s + \gamma_s \bm{\hat w}^* \cdot \bm{U}_s}{\Delta_s}, \label{eqn:alpha_eqs} \\
    \beta_s &= \frac{\tilde \gamma_s  \bm{\hat w} \cdot \bm{U}_s +(1 - \delta_s) \bm{\hat w}^* \cdot \bm{U}_s}{\Delta_s} ,\label{eqn:beta_eqs} 
\end{align}
where
\begin{align}
    \delta_s &= \bm{\hat w} \cdot \left[\frac{\bvec}{\cycls} \times (\bm{\hat w} \cdot \nabla \vzeros) \right] , \label{eqn:coef1} \\
    \tilde \delta_s &= \bm{\hat w}^* \cdot \left[\frac{\bvec}{\cycls} \times (\bm{\hat w}^* \cdot \nabla \vzeros) \right] , \label{eqn:coef2}\\
    \gamma_s &= \bm{\hat w} \cdot \left[\frac{\bvec}{\cycls} \times (\bm{\hat w}^* \cdot \nabla \vzeros) \right] , \label{eqn:coef3}\\
    \tilde \gamma_s &= \bm{\hat w}^* \cdot \left[\frac{\bvec}{\cycls} \times (\bm{\hat w} \cdot \nabla \vzeros) \right] , \label{eqn:coef4}
\end{align}
and the determinant $\Delta_s$ is given by
\begin{align}\label{eqn:determinant}
    \Delta_s = 1 - (\delta_s + \tilde \delta_s) +(\delta_s \tilde \delta_s - \gamma_s \tilde \gamma_s) .
\end{align}
Using the cyclical property of the triple product $\bm{a}_1 \cdot (\bm{a}_2 \times \bm{a}_3) = \bm{a}_2 \cdot (\bm{a}_3 \times \bm{a}_1)$ and $(\bm{\hat w}^*)^* = - \bm{\hat w}$, the coefficients in Eqs~\ref{eqn:coef1}-\ref{eqn:coef4} can be written in more transparent form as
\begin{align}
    \delta_s &= - \bm{\hat w}^* \cdot \left(\frac{\bvec \times \nabla \vzeros}{\cycls}  \right) \cdot \bm{\hat w}^* ,\\
    \tilde \delta_s &= - \bm{\hat w} \cdot \left(\frac{\bvec \times \nabla \vzeros}{\cycls}  \right) \cdot \bm{\hat w} ,\\
    \gamma_s &= \bm{\hat w} \cdot \left(\frac{\bvec \times \nabla \vzeros}{\cycls}  \right) \cdot \bm{\hat w}^* ,\\
    \tilde \gamma_s &= \bm{\hat w}^* \cdot \left(\frac{\bvec \times \nabla \vzeros}{\cycls}  \right) \cdot \bm{\hat w} .
\end{align}
 Substituting the above expressions for the coefficients $\delta_s, \tilde \delta_s, \gamma_s$ and $ \tilde \gamma_s$ into Eq~\ref{eqn:alpha_eqs},~\ref{eqn:beta_eqs} and~\ref{eqn:determinant}, we obtain 
 \begin{align}
     \alpha_s &= \frac{1}{\Delta_s}\left( \bm{\hat w} \cdot \bm{U}_s + \left(\bm{\hat w} \cdot \bm{T}_s \cdot \bm{\hat w}\right) \bm{\hat w} \cdot \bm{U}_s  +  \left(\bm{\hat w} \cdot \bm{T}_s \cdot \bm{\hat w}^*\right) \bm{\hat w}^* \cdot \bm{U}_s \right) ,\label{eqn:alpha_covariant} \\
     \beta_s &= \frac{1}{\Delta_s}\left(  \bm{\hat w}^* \cdot \bm{U}_s + \left(\bm{\hat w}^* \cdot \bm{T}_s \cdot \bm{\hat w}^*\right) \bm{\hat w}^* \cdot \bm{U}_s  +  \left(\bm{\hat w}^* \cdot \bm{T}_s \cdot \bm{\hat w}\right) \bm{\hat w} \cdot \bm{U}_s\right) ,\label{eqn:beta_covariant} \\
     \Delta_s &=1 + (\bm{\hat w} \cdot \bm{T}_s \cdot \bm{\hat w} + \bm{\hat w}^* \cdot \bm{T}_s \cdot \bm{\hat w}^*) \nonumber \\
     &\quad + [(\bm{\hat w} \cdot \bm{T}_s \cdot \bm{\hat w})( \bm{\hat w}^* \cdot \bm{T}_s \cdot \bm{\hat w}^*) - (\bm{\hat w} \cdot \bm{T}_s \cdot \bm{\hat w}^*)(\bm{\hat w}^* \cdot \bm{T}_s \cdot \bm{\hat w}) ] ,\label{eqn:determinant_covariant}
 \end{align}
 with $\bm{T}_s \equiv \left(\bvec/\cycls \times \nabla \vzeros  \right)_\perp : W_\perp \to W_\perp$ the linear operator $\bvec/\cycls \times \nabla \vzeros$ restricted to act on the subspace perpendicular to $\bvec(\bm{x},t)$ at position $\bm{x}$ and time $t$, defined by $W_\perp \equiv \{\bm{y} \in \mathbb{R}^3 | \bvec(\bm{x},t) \cdot \bm{y} = 0\}$.
 Substituting Eq~\ref{eqn:alpha_covariant} and~\ref{eqn:beta_covariant} into Eq~\ref{eqn:vpol_proj}, and recalling that $\bm{U}_s = \bm{\hat w}(\bm{\hat w} \cdot \bm{U}_s) + \bm{w}^*(\bm{\hat w}^* \cdot \bm{U}_s) $, we obtain an expression for $\vpols$
\begin{align}\label{eqn:vec_Gpol_result}
    \vpols = \frac{1 }{\Delta_s}(\bm{U}_s + [\bm{e}^i (\bm{e}_i \cdot \bm{T}_s \cdot \bm{e}^j) \bm{e}_j] \cdot \bm{U}_s) ,
\end{align}
with $\bm{e}^i = \{\bm{\hat w}, \bm{\hat w}^* \}$ the orthonormal basis elements of $W_\perp$. Since $\bm{\hat w}$ has arbitrary direction in the plane perpendicular to $\bvec$, we can express the result in a basis-independent form as
\begin{align}
     \vpols = \frac{1 +\bm{T}_s }{\Delta_s} \cdot \bm{U}_s = \frac{1 +(\bvec \times \nabla \vzeros)_\perp / \cycls}{\Delta_s} \cdot \bm{U}_s .
\end{align}
We note that $\bm{U}_s$ in Eq~\ref{eqn:R_def} is an element of the orthogonal subspace $W_\perp$ and therefore, $(\bvec \times \nabla \vzeros)_\perp \cdot \bm{U}_s = (\bvec \times \nabla \vzeros) \cdot \bm{U}_s$. The determinant $\Delta_s$ in basis-independent form is constructed from Eq~\ref{eqn:determinant_covariant} and can be written as
\begin{align}\label{eqn:determinant_expr}
    \Delta_s = 1 + \text{tr}(\bm{T}_s) + \text{det}(\bm{T}_s) ,
\end{align}
or, explicitly, using the definition of $\bm{T}_s$, as
\begin{align}\label{eqn:det_partial}
    \Delta_s = 1 + \text{tr}\left( \left[\frac{\bvec}{\cycls} \times \nabla \vzeros\right]_\perp \right) + \text{det}\left(\left[\frac{\bvec}{\cycls} \times \nabla \vzeros\right]_\perp\right).
\end{align}
The trace term is given by 
\begin{align}\label{eqn:trace_levi}
    \text{tr}\left( \left[\frac{\bvec}{\cycls} \times \nabla \vzeros\right]_\perp \right) = \frac{1}{\cycls}\epsilon_i^{\,\, jk}b_k \nabla_j \bar v_s^i = \frac{1}{\cycls}\bvec \cdot (\nabla \times \vzeros).
\end{align} 
Meanwhile, for any linear application $A : \mathbb{R}^3 \to \mathbb{R}^3$, the determinant of its restriction to the $W_\perp$ subspace can be expressed as $\text{det}(A_\perp) =   \epsilon_{\perp}^{kl} \epsilon_{\perp \,\, ij} A^i_{\,\, k} A^{j}_{\,\,l}/2$, where $\epsilon_\perp^{ij} \equiv \epsilon^{ijk}b_k$ denotes the induced Levi-Civita symbol on  $W_\perp$. The determinant term in Eq~\ref{eqn:det_partial} becomes
\begin{align}\label{eqn:determinant_levi}
    \text{det}\left(\left[\frac{\bvec}{\cycls} \times \nabla \vzeros\right]_\perp\right) &= \frac{1}{2} \epsilon_{\perp}^{kl}\epsilon_{\perp ij}\left(\frac{\bvec}{\cycls} \times \nabla \vzeros \right)^i_{\,\, k}  \left(\frac{\bvec}{\cycls} \times \nabla \vzeros \right)^j_{\,\, l}  \nonumber \\
    &= \frac{1}{2 \cycls^2} \epsilon_{\perp}^{kl} \epsilon_{\perp \, i j} (\nabla \vzeros)^i_{\,\, k}  (\nabla \vzeros)^j_{\,\, l} = \frac{1}{\cycls^2}\text{det} ( \nabla \vzeros )_\perp ,
\end{align}
where we used the fact that $(\bvec \times \nabla \vzeros)^i_{\,\, k} = \epsilon_\perp^{ij} \nabla_j \bar v_k$ and $\epsilon_{\perp}^{ki} \epsilon_{\perp \, kj} = (\Pi_\perp)^i_{\,\, j}$. Finally, the determinant $\Delta_s$ in Eq~\ref{eqn:determinant_expr} becomes 
\begin{equation}\label{eqn:determinant_expanded}
    \Delta_s = 1 + \frac{\bvec \cdot (\nabla \times \vzeros)}{\cycls} + \frac{\text{det} ( \nabla \vzeros )_\perp}{\cycls^2} , 
\end{equation}
which contains a correction due to the parallel component of the vorticity $\bvec \cdot (\nabla \times \vzeros)/\cycls \sim O(\epsilon)$ and a nonlinear correction related to flow shear $\text{det} ( \nabla \vzeros )_\perp/\cycls^2 \sim O(\epsilon^2)$. For a linear geometry in the cold-ion approximation, and for the case where the leading-order ion perpendicular flow is purely the $E \times B$ drift $\vperpzero = \vexb$, Eq~\ref{eqn:determinant_expanded} reduces to Eq (A6) in~\cite{reiserAIP_2012}. 
The final solution for the polarisation velocity is therefore given by Eq~\ref{eqn:vpol_closure}, or in more succinct form, by 
\begin{align}\label{eqn:Gpol_solution}
    \vpols = \bm{Q}_s(\vzeros) \cdot  \bm{U}_s,
\end{align}
where
\begin{equation}\label{eqn:Qs_cons}
    \bm{Q}_s(\vzeros) = \frac{1 + (\bvec \times \nabla \vzeros)/\cycls }{1 + \bvec \cdot (\nabla \times \vzeros)/\cycls + \text{det}(\nabla \vzeros)_\perp/\cycls^2}.
\end{equation}
Within the drift-ordering $\vzeros \cdot \nabla/\cycls \sim \epsilon$, the application $\bm{Q}_s^{-1}$ is invertible, as the determinant $\Delta_s = 1 + O(\epsilon)$ does not vanish. In App~\ref{sec:solution}, we verify that the expression of $\vpols$ given in Eq~\ref{eqn:Gpol_solution} satisfies Eq~\ref{eqn:Gpol_start}. 

We note that Eq~\ref{eqn:Gpol_solution} represents a non-perturbative result, containing terms of all orders in $\epsilon$, as can be seen by Taylor-expanding the denominator in Eq~\ref{eqn:Qs_cons}.
All of the subleading terms contained in $\bm{Q}_s$ are required for the existence of exact conservation laws.

\section{Conservative drift-reduced model}\label{sec:conservation_proof}

Based on Eq~\ref{eqn:vpol_closure} we construct now the drift-reduced model, valid to $O(\epsilon)$, which admits energy and momentum conservation laws satisfied exactly, i.e., to all orders in $\epsilon$. The model evolves the density, parallel velocity and pressure of each species and is closed by a vorticity equation, an associated Poisson equation for $\phi$, and Ampère's equation for the vector potential $\bm{A}$. 

Expressing the total fluid velocity as $\bm{V}_s = \vzeros + \vpols $ in Eqs~\ref{eqn:continuity} and~\ref{eqn:pressure}, we obtain
\begin{align}
    &\frac{\partial n_s}{\partial t} + \nabla \cdot [n_s (\vzeros + \vpols)] = \mathcal{S}_s , \label{eqn:continuity_no_indx} \\
    &\frac{\partial }{\partial t}\left( \frac{3}{2} p_s \right) + \nabla \cdot \left[\frac{3}{2}p_s (\vzeros + \vpols) + \bm{q}_s \right] + \bm{P}_s : \nabla (\vzeros + \vpols) = \mathcal{Q}_s +r_s \mathcal{\overline{ K}}_s \label{eqn:pressure_no_indx},
\end{align}
where $\vpols$ is either given implicitly by Eq~\ref{eqn:vpol_closure}, or explicitly via Eq~\ref{eqn:vpol_solution_full_explicit}, and we define $\mathcal{\overline{ K}}_s \equiv m_s n_s \bar v_s^2/2$ to be the leading-order component of the fluid kinetic energy. The equation for the velocity component parallel to the magnetic field $v_{\parallel s} = \bm{V}_s \cdot \bvec$ of species $s$, is obtained by projecting the $O(\epsilon)$ expanded momentum equation~\ref{eqn:momentum}
\begin{equation}
    \frac{\partial(m_s n_s \vzeros)}{\partial t} + \nabla \cdot (m_s n_s \bm{V}_s \vzeros) + \nabla \cdot \bm{P}_s - q_s n_s(\bm{E} + \bm{V}_s \times \bm{B}) = \bm{\mathcal{R}}_s,
\end{equation}
onto $\bvec$, obtaining
\begin{equation}\label{eqn:par_momentum_no_index}
    \frac{\partial (m_s n_s v_{\parallel s})}{\partial t} + \nabla \cdot (m_s n_s v_{\parallel s} \bm{V}_s) - m_s n_s \vzeros \cdot \frac{d_s \bvec}{dt } +  (\nabla \cdot  \bm{P}_s)_\parallel = \mathcal{R}_{\parallel s} + q_s n_s E_{\parallel},
\end{equation}
 with $(\nabla \cdot \bm{P}_s)_\parallel \equiv \bvec \cdot (\nabla \cdot \bm{P}_s)$ and, similarly, for other quantities.

To derive the vorticity equation, we impose the quasi-neutrality constraint $\nabla \cdot \bm{J} = 0$. 
We begin with the force-balance equation for the leading-order component of the total momentum density, $\Mtot = \sum_s m_s n_s \vzeros$. As shown in Sec~\ref{sec:conservation_properties}, this is given by
\begin{align}\label{eqn:LO_momentum_density_first_time} 
     \frac{\partial \Mtot}{\partial t} + \nabla \cdot \sum_s ( \bm{V}_s \Mtot_s + \bm{P}_s)  &= \bm{J} \times \bm{B} + \sum_s \bm{\mathcal{R}}_s ,
\end{align}
where the total current $\bm{J} = \Jpar + \Jperpzero + \Jpol$ is the sum of the leading-order components, $\Jpar + \Jperpzero = \sum_s q_s n_s \vpars + \sum_s q_s n_s \vperpzeros$, and the subleading polarisation term, $\Jpol = \sum_s q_s n_s \vpols$. 
We note the similarity between Eq~\ref{eqn:LO_momentum_density_first_time} with the non drift-reduced form in Eq~\ref{eqn:total_momentum_density}.
Taking the cross-product of Eq~\ref{eqn:LO_momentum_density_first_time} with $\bm{B}$ leads to
\begin{equation}
    \bm{J} = \Jpar + \frac{\bvec}{B}\times \frac{\partial \Mtot}{\partial t} + \frac{\bvec}{B}\times \left(\sum_s[\nabla \cdot  ( \bm{V}_s \Mtot_s + \bm{P}_s) - \bm{\mathcal{R}}_s ] \right), 
\end{equation}
which can be rewritten as
\begin{equation}
    \bm{J} = \Jpar + \frac{\partial}{\partial t}\left(\frac{\bvec \times \Mtot}{B}\right)  + \frac{\bvec}{B}\times \left(\sum_s[\nabla \cdot  ( \bm{V}_s \Mtot_s + \bm{P}_s) - \bm{\mathcal{R}}_s ] \right) - \frac{\partial}{\partial t}\left(\frac{\bvec}{B}\right)\times \Mtot.
\end{equation}
Imposing the quasi-neutrality constraint $\nabla \cdot \bm{J} = 0$, we obtain the vorticity equation for $\varpi \equiv -\nabla \cdot (\bvec \times \Mtot/B)$, that is
\begin{equation}\label{eqn:scalar_vorticity_new}
    \frac{\partial \varpi}{\partial t} = \nabla \cdot (\Jpar + \Jperpzero)  +  \nabla \cdot \left[\sum_s \frac{\bvec}{B}\times \nabla \cdot  ( \bm{V}_s \Mtot_s) \right] +   \nabla \cdot \left[ \left( \frac{(\bm{\Pi}_\perp - \bvec \bvec)\cdot \nabla \times \bm{E}}{B^2}\right) \times \Mtot \right],
\end{equation}
where we use Faraday's law and, since $\bvec \cdot \partial_t \bvec = 0$, $\partial_t(\bvec/B) = 2 \bvec \bvec \cdot (\nabla \times \bm{E}) /B^2 -  \nabla \times \bm{E}/B^2 = (\bvec \bvec  - \bm{\Pi}_\perp) \cdot (\nabla \times \bm{E}) /B^2$, with $\bm{\Pi}_\perp = \mathbb{I} - \bvec \bvec$ the perpendicular projection operator. In Eq~\ref{eqn:scalar_vorticity_new}, we also express the pressure and momentum drive terms as $ \sum_s \bvec \times (\nabla \cdot \bm{P}_s - \bm{\mathcal{R}}_s)/B  = \sum_s q_s n_s \vperpzeros = \Jperpzero$. Using $\bm{V}_s = \vzeros + \vpols$ in Eq~\ref{eqn:scalar_vorticity_new}, we obtain
\begin{align}\label{eqn:scalar_vorticity_final}
    \frac{\partial \varpi}{\partial t} &= \nabla \cdot (\Jpar + \Jperpzero)  + \nabla \cdot \left[\sum_s \frac{\bvec}{B}\times \nabla \cdot  ( \vzeros \Mtot_s) \right] +   \nabla \cdot \left[ \left( \frac{(\bm{\Pi}_\perp - \bvec \bvec)\cdot \nabla \times \bm{E}}{B^2}\right) \times \Mtot \right] \nonumber \\
    &
    + \nabla \cdot \left[\sum_s \frac{\bvec}{B}\times \nabla \cdot  ( \bm{Q}_s(\vzeros) \cdot \bm{U}_s \Mtot_s) \right],
\end{align}
where $\vpols$ is expressed in terms of leading-order quantities via Eq~\ref{eqn:Gpol_solution}, with $\bm{U}_s$ given in Eq~\ref{eqn:R_def}. The last term in Eq~\ref{eqn:scalar_vorticity_final} is the conservative correction to the usual vorticity equation found in the literature, accounting for polarisation advection of the leading-order momentum $\Mtot$. Having evolved the scalar vorticity in time, the Poisson equation, defined via $\varpi = -\nabla \cdot (\bvec \times \Mtot/B)$, takes its usual form(see, e.g.,~\citet{Giacomin_2022}),
\begin{equation}\label{eqn:Poisson_new}
    \nabla \cdot \left( \frac{\rho}{B^2}\nabla_\perp \phi  \right) = \varpi - \nabla \cdot \left(\sum_s \frac{(\nabla \cdot \bm{P}_s)_\perp - \bm{\mathcal{R}}_{\perp s} }{B \cycls}\right) - \nabla \cdot \left( \frac{\rho}{B^2}\left.\frac{\partial \bm{A}}{\partial t}\right|_\perp\right),
\end{equation}
where the terms due to $\bm{\mathcal{R}}_{\perp s}$ and $\partial_t \bm{A}|_\perp$ are usually neglected because of additional ordering assumptions on the size of perpendicular friction terms and strength of electromagnetic effects~\citep{zeilerdrakerogers_DR_1997}. In Eq~\ref{eqn:Poisson_new}, $\rho \equiv \sum_s m_s n_s$ is the total mass density of the plasma and the perpendicular component of a given vector $\bm{W}$ is defined using the projection operator, $\bm{W}_\perp \equiv \bm{\Pi}_\perp \cdot \bm{W}$.

To summarise, the drift-reduced fluid model we propose, obeying exact conservation laws for the leading-order energy $\mathcal{\overline{ H}}$, leading-order momentum $\Mtot$, mass and charge (as shown in Sec~\ref{sec:conservation_properties}), is given by the moment equations Eqs~\ref{eqn:continuity_no_indx}, \ref{eqn:pressure_no_indx} and~\ref{eqn:par_momentum_no_index}, the vorticity equation Eq~\ref{eqn:scalar_vorticity_final} for $\varpi$, the Poisson equation Eq~\ref{eqn:Poisson_new} for $\phi$, the polarisation drift expression Eq~\ref{eqn:vpol_solution_full_explicit} for $\vpols$, and the Ampère equation for $\bm{A}$, which in the Coulomb gauge, $\nabla \cdot \bm{A} = 0$, is given by
\begin{align}\label{eqn:ampere_DR}
    \nabla^2 \bm{A} = - \mu_0 (\Jpar + \Jperpzero + \Jpol).
\end{align}
Computing the polarisation velocity requires evaluating $\partial_t \vzeros$ in Eq~\ref{eqn:vpol_solution_full_explicit}, which is itself a function of $\vpols$.  Solution of this coupled system can be either implemented implicitly via an iterative procedure, or explicitly by using $\vpols$ from the previous time-step to first evolve the dynamical fields, then calculating $\partial_t \vzeros$ to update $\vpols$. Devising a numerical implementation which preserves the conservation properties of the system is however left for future work.
The commonly-used non-conservative drift-reduced vorticity equation is obtained from Eq~\ref{eqn:scalar_vorticity_final} by keeping only the leading-order terms. This amounts to neglecting the last term in Eq~\ref{eqn:scalar_vorticity_final}, yielding
\begin{equation}\label{eqn:noncons_scalar_vorticity}
    \frac{\partial \varpi}{\partial t} \simeq \nabla \cdot (\Jpar + \Jperpzero)  + \nabla \cdot \left[\sum_s \frac{\bvec}{B}\times \nabla \cdot  ( \vzeros \Mtot_s) \right] +   \nabla \cdot \left[ \left( \frac{(\bm{\Pi}_\perp - \bvec \bvec)\cdot \nabla \times \bm{E}}{B^2}\right) \times \Mtot \right],
\end{equation}
which is the form found in the literature and implemented in drift-reduced turbulence codes~\citep{Giacomin_2022, Dull_2024, gath_wiesenberger_AIP_2019, Stegmeir_2019}. For the case considered in~\citet{reiserAIP_2012}, the use of a conservative formulation had a small influence on the overall turbulent dynamics. In fact, the corrections needed to ensure energy-momentum consistency are subdominant in the drift-ordering expansion. However, in regimes where flow shear becomes large, and the tensor $\bm{Q}_s$ in Eq~\ref{eqn:Qs_cons} deviates sufficiently from unity, a self-consistent treatment of the vorticity equation, as given in Eq~\ref{eqn:scalar_vorticity_final}, could become necessary to accurately describe the nonlinear dynamics.

\section{Conservation properties}\label{sec:conservation_properties}
To prove that the system of equations Eqs~\ref{eqn:continuity_no_indx}, \ref{eqn:pressure_no_indx}, \ref{eqn:par_momentum_no_index}, \ref{eqn:scalar_vorticity_final} and \ref{eqn:Poisson_new} conserves both energy and momentum, we follow the procedure in~\cite{halpern_energy2023}. To obtain the transport equation for the parallel kinetic energy density $\mathcal{K}_{\parallel s} \equiv m_s n_s v_{\parallel s}^2/2$ of species $s$, we multiply Eq~\ref{eqn:par_momentum_no_index} by $v_{\parallel s}$. After straightforward algebra, this leads to
\begin{align}\label{eqn:par_energy}
    \frac{\partial \mathcal{K}_{\parallel s}}{\partial t} + \nabla \cdot \left(\mathcal{K}_{\parallel s}\bm{V}_s\right) - m_s n_s v_{\parallel s} \vzeros \cdot \frac{d_s \bvec}{dt}   &+ \vpars \cdot (\nabla \cdot  \bm{P}_s) =  \vpars \cdot \bm{\mathcal{R}}_s + q_s n_s \vpars \cdot \bm{E} - r_s \mathcal{K}_{\parallel s}.
\end{align}
We note that $\bvec \cdot d_s \bvec/dt  = d_s (b^2/2)/dt= 0$ and, as a consequence, $\vzero \cdot d_s \bvec /dt = \vperpzeros \cdot d_s\bvec/dt$.
Adding the internal energy $\mathcal{U}_s = 3 p_s/2$ contribution in~\ref{eqn:pressure_no_indx} to Eq~\ref{eqn:par_energy}, we have 
\begin{align}\label{eqn:par_internal_energy}
    \frac{\partial (\mathcal{K}_{\parallel s} + \mathcal{U}_s) }{\partial t} + \nabla \cdot \left[(\mathcal{K}_{\parallel s} + \mathcal{U}_s)\bm{V}_s + \bm{q}_s\right] &- m_s n_s v_{\parallel s} \vperpzeros \cdot \frac{d_s \bvec}{dt} + \bm{P}_s: \nabla \bm{V}_s + \vpars \cdot (\nabla \cdot  \bm{P}_s) \nonumber \\
    &= \mathcal{Q}_s +  \vpars \cdot \bm{\mathcal{R}}_s + q_s n_s \vpars \cdot \bm{E} + r_s (\mathcal{\overline{ K}}_{s} -\mathcal{K}_{\parallel s}).
\end{align}
To find the transport equation for the leading-order perpendicular kinetic energy $\mathcal{\overline{ K}}_{\perp s} \equiv m_s n_s \bar v_{\perp s}^2/2 = \mathcal{\overline{ K}}_{s} -\mathcal{K}_{\parallel s}$ and the magnetic field energy $\mathcal{H}_B$, we multiply the quasi-neutrality equation $\nabla \cdot \bm{J} = 0$ by $\phi$ and express this as
\begin{align}\label{eqn:qn_energy_integral}
    0 = \nabla \cdot ( \phi \bm{J} ) - (\Jpar + \Jperpzero + \Jpol) \cdot \nabla \phi,
\end{align}
where we recall $\bm{J} \equiv \sum_s q_s n_s \bm{V}_s = \Jpar + \Jperpzero + \Jpol$. The total polarisation current can be expressed as $\Jpol \equiv \sum_s \Jpols = \sum_s q_s n_s \vpols $, with $\vpols$ given by Eq~\ref{eqn:vpol_closure}. The polarisation current density of species $s$, $\Jpols$, is therefore
\begin{align}\label{eqn:Jpol0}
    \Jpols =   \frac{m_s n_s}{B} \bvec\times \left(\frac{d_s  \vzeros}{d t} + r_s \vzeros \right) .
\end{align}  
As a first step, we note that, since the polarisation current is orthogonal to $\bvec$, we can express
\begin{align}\label{eqn:Jpol_0}
    \Jpol \cdot \nabla \phi = \Jpol \cdot \nabla_\perp \phi = - \Jpol \cdot [\bvec \times (\bvec \times \nabla \phi)], 
\end{align}
and, given that $ \nabla \phi = -\bm{E} - \partial_t \bm{A}$, we can rewrite Eq~\ref{eqn:Jpol_0} as
\begin{align}
    \Jpol \cdot \nabla \phi = -  \Jpol \cdot (\bm{B} \times \vexb) -  \Jpol \cdot \partial_t \bm{A}.
\end{align}
Moreover, since $\vexb$ is species-independent, we obtain
\begin{equation}\label{eqn:part_jpol}
    \Jpol \cdot \nabla \phi = - \sum_s  \Jpols \cdot (\bm{B} \times \vexb) -  \Jpol \cdot \partial_t \bm{A}.
\end{equation}
Expressing the $E \times B$ drift in Eq~\ref{eqn:part_jpol} as $\vexb = \vperpzeros - \bm{w}_{\perp s}$, with $\bm{w}_{\perp s}$ the drift-velocity of species $s$ in the frame comoving with the $E \times B$ velocity (cf. Eq~\ref{eqn:leading_order_vperp_decomp}), we have
\begin{align}\label{eqn:Jpol1}
    \sum_s \Jpols  \cdot (\bm{B} \times \vexb) = \sum_s \left[m_s n_s \left(\frac{d_s \vzeros}{dt} + r_s \vzeros \right) \cdot \vperpzeros \right] - \sum_s  \Jpols \cdot (\bm{B} \times \bm{w}_{\perp s}).
\end{align}
Decomposing $\vzeros = v_{\parallel s} \bvec + \vperpzeros$ into its parallel and perpendicular components, the first term on the right-hand-side of Eq~\ref{eqn:Jpol1} becomes
\begin{align}\label{eqn:Jpol2}
    \sum_s \left[m_s n_s \left(\frac{d_s \vzeros}{dt} + r_s \vzeros \right) \cdot \vperpzeros \right] &=   \sum_s m_s n_s   \left(\vperpzeros \cdot \frac{d_s  \vperpzeros}{d t} + r_s \bar v_{\perp s}^2\right) \nonumber \\
    &+ \sum_s m_s n_s v_{\parallel s} \vperpzeros \cdot \frac{d_s \bvec}{dt} .
\end{align}
The first term on the right-hand-side of Eq~\ref{eqn:Jpol2} leads to a transport law for the leading-order perpendicular kinetic energy of species $s$,  $ \mathcal{\overline{ K}}_{\perp s}$. Indeed, we have
\begin{align}\label{eqn:Jpol3}
     m_s n_s \vperpzeros \cdot \left(\frac{d_s  \vperpzeros}{d t} + r_s \vperpzeros \right) &= \frac{\partial \mathcal{\overline{ K}}_{\perp s}}{\partial t} + \nabla \cdot (\mathcal{\overline{ K}}_{\perp s} \bm{V}_s) + r_s \mathcal{\overline{ K}}_{\perp s},
\end{align}
where the continuity equation $d_s n_s/ dt = r_s n_s - n_s \nabla \cdot \bm{V}_s$ is used.
Using Eqs~\ref{eqn:part_jpol}, \ref{eqn:Jpol1}, \ref{eqn:Jpol2} and \ref{eqn:Jpol3} we find
\begin{align}\label{eqn:Jpol_dot_nabla_phi}
    \Jpol \cdot \nabla \phi =  &-\sum_s  \left[\frac{\partial \mathcal{\overline{ K}}_{\perp s}}{\partial t} + \nabla \cdot (\mathcal{\overline{ K}}_{\perp s} \bm{V}_s) + r_s \mathcal{\overline{ K}}_{\perp s} \right] \nonumber \\
    &- \sum_s m_s n_s v_{\parallel s} \vperpzeros \cdot \frac{d_s \bvec}{dt} + \sum_s \Jpols \cdot (\bm{B} \times \bm{w}_{\perp s}) - \Jpol \cdot  \partial_t \bm{A}. 
\end{align}
Given Eq~\ref{eqn:Jpol_dot_nabla_phi}, we can rewrite Eq~\ref{eqn:qn_energy_integral} as
\begin{align}\label{eqn:perp_energy_transp1}
     &\nabla \cdot ( \phi \bm{J} ) - (\Jpar + \Jperpzero) \cdot \nabla \phi - \sum_s \Jpols \cdot (\bm{B} \times \bm{w}_{\perp s}) + \Jpol \cdot  \partial_t \bm{A} \nonumber \\
    &+\sum_s  \left[\frac{\partial \mathcal{\overline{ K}}_{\perp s}}{\partial t} + \nabla \cdot (\mathcal{\overline{ K}}_{\perp s} \bm{V}_s) + r_s \mathcal{\overline{ K}}_{\perp s} \right] + \sum_s m_s n_s v_{\parallel s} \vperpzeros \cdot \frac{d_s \bvec}{dt} = 0.
\end{align}
The term $\nabla \cdot (\phi \bm{J})$ can be expressed in terms of the Poynting flux as
\begin{align}
    \nabla \cdot (\phi \bm{J}) = \nabla \cdot \bm{S} + \nabla \cdot \left(\frac{\partial_t \bm{A} \times \bm{B}}{\mu_0} \right),
\end{align}
where $\nabla \times \bm{B} = \mu_0 \bm{J}$ is used. Writing the term $\nabla \cdot (\partial_t \bm{A} \times \bm{B}/\mu_0)$ as
\begin{align}
    \nabla \cdot \left(\frac{\partial_t \bm{A} \times \bm{B}}{\mu_0} \right) = \frac{\bm{B}}{\mu_0} \cdot \partial_t \nabla \times \bm{A}  -  \frac{\nabla \times \bm{B}}{\mu_0} \cdot \partial_t \bm{A}= \frac{\partial \mathcal{H}_B}{\partial t} - \bm{J} \cdot \partial_t \bm{A},
\end{align}
where we recall that $\mathcal{H}_B = B^2/(2 \mu_0)$ is the magnetic energy density, we obtain
\begin{align}\label{eqn:div_phi_J}
    \nabla \cdot (\phi \bm{J}) = \frac{\partial \mathcal{H}_B}{\partial t} + \nabla \cdot \bm{S} - \bm{J} \cdot \partial_t \bm{A} .
\end{align}
Substituting Eq~\ref{eqn:div_phi_J} into Eq~\ref{eqn:perp_energy_transp1}, we find the transport equation for the sum of the field energy, $\mathcal{H}_B$, and the total leading-order perpendicular kinetic energy, $\mathcal{\overline{ K}}_\perp = \sum_s \mathcal{\overline{ K}}_{\perp s}$, given by
\begin{align}\label{eqn:magnetic_perp_energy}
     &\frac{\partial (\mathcal{H}_B + \mathcal{\overline{ K}}_\perp)}{\partial t} + \nabla \cdot (\bm{S} + \sum_s \mathcal{\overline{ K}}_{\perp s} \bm{V}_s) + (\Jpar + \Jperpzero) \cdot \bm{E}   \nonumber \\
    & + \sum_s r_s \mathcal{\overline{ K}}_{\perp s} + \sum_s m_s n_s v_{\parallel s} \vperpzeros \cdot \frac{d_s \bvec}{dt}  - \sum_s \Jpols \cdot (\bm{B} \times \bm{w}_{\perp s}) = 0 .
\end{align}
We now add the contribution from the parallel and internal energy densities in Eq~\ref{eqn:par_internal_energy}, having summed it over the species indices $s$, to Eq~\ref{eqn:magnetic_perp_energy}. We thus obtain that the total leading-order energy density, $\mathcal{\overline{ H}} \equiv \mathcal{H}_B + \sum_s(\mathcal{\overline{ K}}_s + \mathcal{U}_s)$, evolves according to
\begin{align}\label{eqn:total_energy_transp1}
    \frac{\partial \mathcal{\overline{ H}} }{\partial t} &+ \nabla \cdot \left(\bm{S} + \sum_s [\mathcal{\overline{ H}}_s\bm{V}_s + \bm{q}_s]\right) + \sum_s \bm{P}_s : \nabla \bm{V}_s + \sum_s \vpars \cdot (\nabla \cdot \bm{P}_s) \nonumber \\
    &+ (\Jpar + \Jperpzero) \cdot \bm{E}    - \sum_s q_s n_s \vpars \cdot \bm{E} - \sum_s \Jpols \cdot (\bm{B} \times \bm{w}_{\perp s}) \nonumber \\
    &\qquad \qquad = \sum_s (\mathcal{Q}_s  + \vpars \cdot \bm{\mathcal{R}}_s ) , 
\end{align}
where we introduce the species specific energy density, $\mathcal{\overline{ H}}_s \equiv \mathcal{\overline{ K}}_s + \mathcal{U}_s$.
 Finally, we rewrite the term $\sum_s \Jpols \cdot (\bm{B} \times \bm{w}_{\perp s})$ by decomposing $\vpols = \vperps - \vperpzeros = \vperps - \vexb - \bm{w}_{\perp s} $, that is
\begin{align}\label{eqn:Jpol_cdot_B_times_wperp}
    \Jpols \cdot (\bm{B} \times \bm{w}_{\perp s}) = q_s n_s (\vperps - \vexb - \bm{w}_{\perp s}) \cdot (\bm{B} \times \bm{w}_{\perp s})  = q_s n_s (\vperps - \vexb) \cdot (\bm{B} \times \bm{w}_{\perp s}). 
\end{align}
Given that $\vexb = \bm{E} \times \bm{B}/B^2$, one deduces that the term $ qn \vexb \cdot (\bm{B} \times \bm{w}_{\perp s}) = -q_s n_s \bm{w}_{\perp s} \cdot \bm{E}$. Furthermore, we have 
\begin{align}\label{eqn:B_times_wperp}
    q_s n_s (\bm{B} \times \bm{w}_{\perp s}) = \bm{B} \times (\vdiams +\vpis + \vRs) = - (\nabla \cdot \bm{P}_s)_\perp + \bm{\mathcal{R}}_{\perp s}. 
\end{align} 
Substituting the expression Eq~\ref{eqn:B_times_wperp} into Eq~\ref{eqn:Jpol_cdot_B_times_wperp} we obtain
\begin{align}
    \Jpols \cdot (\bm{B} \times \bm{w}_{\perp s}) = - \vperps \cdot  (\nabla \cdot \bm{P}_s) + \vperps \cdot  \bm{\mathcal{R}}_s   + q_s n_s \bm{w}_{\perp s} \cdot \bm{E}, 
\end{align}
which, substituted into Eq~\ref{eqn:total_energy_transp1}, yields
\begin{align}\label{eqn:total_energy_transp2}
    \frac{\partial \mathcal{\overline{ H}} }{\partial t} &+ \nabla \cdot \left(\bm{S} + \sum_s [\mathcal{\overline{ H}}_s \bm{V}_s + \bm{P}_s \cdot \bm{V}_s + \bm{q}_s]\right) \nonumber \\
    &+ (\Jpar + \Jperpzero) \cdot \bm{E} - \sum_s q_s n_s (\vpars + \bm{w}_{\perp s} ) \cdot \bm{E} = \sum_s(\mathcal{Q}_s  + \bm{V}_s \cdot \bm{\mathcal{R}}_s ) .
\end{align} 
 Recalling that
 \begin{equation}
     \Jpar + \Jperpzero = \sum_s q_s n_s (\bm{v}_{\parallel s} + \bm{w}_{\perp s}),
 \end{equation} 
 we find that the ohmic heating terms in Eq~\ref{eqn:total_energy_transp2} cancel, and the leading-order energy transport equation is given by
\begin{align}\label{eqn:LO_energy_transp}
    \frac{\partial \mathcal{\overline{ H}} }{\partial t} + \nabla \cdot \left(\bm{S} + \sum_s [\mathcal{\overline{ H}}_s\bm{V}_s + \bm{P}_s \cdot \bm{V}_s + \bm{q}_s]\right) 
      &= \sum_s (\mathcal{Q}_s  + \bm{V}_s \cdot \bm{\mathcal{R}}_s ) .
\end{align}
Comparing this result to the time-evolution of the total energy density $\mathcal{H}$, given by Eq~\ref{eqn:energy_full_system}, we find that they have identical form. The model given by Eqs~\ref{eqn:continuity_no_indx},\ref{eqn:pressure_no_indx} and \ref{eqn:par_momentum_no_index}, with $\vpols$ given by Eq~\ref{eqn:vpol_closure}, and coupled to the quasi-neutral Maxwell equations, Eqs~\ref{eqn:ampere}-\ref{eqn:quasi_neutral}, therefore conserves the leading-order component $\mathcal{\overline{ H}}$ of the total energy $\mathcal{H} = \mathcal{\overline{ H}}[1 + O(\epsilon)]$ exactly, that is, to all orders in $\epsilon$. This result can be extended to plasmas with finite Debye length, though for simplicity we have assumed $\partial_t / \omega_p = 0$.  

The leading-order component of the total fluid momentum density $\Mtot \equiv \sum_s m_s n_s \vzeros = \bm{\mathcal{M}}_\parallel + \Mtot_\perp$ is also a conserved quantity. The equation for the parallel momentum of species $s$, $ \bm{\mathcal{M}}_{\parallel s} \equiv m_s n_s \vpars$, is obtained by multiplying  Eq~\ref{eqn:par_momentum_no_index} by $\bvec$, that is
\begin{equation}\label{eqn:par_mom1}
    \frac{\partial \bm{\mathcal{M}}_{\parallel s}}{\partial t} +  \nabla \cdot (\bm{V}_s \bm{\mathcal{M}}_{\parallel s}) -  \mathcal{M}_{\parallel s} \frac{d_s \bvec}{dt} - \Mtot_{\perp s} \cdot \frac{d_s \bvec}{dt} \bvec +  (\nabla \cdot \bm{P}_s)_\parallel \bvec - q_s n_s \bm{E}_\parallel = \bm{\mathcal{R}}_{\parallel s},
\end{equation}
where $\Mtot_{\perp s} \equiv m_s n_s \vperpzeros$ is defined. 
The evolution of the perpendicular momentum $\Mtot_{\perp s}$ is obtained from the force balance equation
\begin{equation}\label{eqn:perp_mom1}
    m_s n_s \left.\frac{d_s \vzeros}{dt}\right|_\perp + r_s m_s n_s \vperpzeros - q_s n_s \bm{B} \times \vperpzeros = q_s n_s \bm{V}_s \times \bm{B},
\end{equation}
which is a rewriting of the polarisation velocity definition, Eq~\ref{eqn:vpol_closure}. Using the continuity equation, Eq~\ref{eqn:continuity_no_indx},  Eq~\ref{eqn:perp_mom1} can be recast as
\begin{equation}\label{eqn:perp_mom2}
    \left.\frac{d_s \Mtot_s}{d t}\right|_\perp + \Mtot_{\perp s}  \nabla \cdot \bm{V}_s - q_s n_s \bm{B} \times \vperpzeros = q_s n_s \bm{V}_s \times \bm{B}.
\end{equation}
Using the fact that
\begin{equation}
    \left.\frac{d_s \Mtot_s}{d t}\right|_\perp = \frac{d_s \Mtot_{\perp s}}{d t} + \Mtot_{\perp s} \cdot \frac{d_s \bvec}{dt} \bvec + \mathcal{M}_{\parallel s} \frac{d_s \bvec}{dt}, 
\end{equation}
we have, upon summing together Eq~\ref{eqn:par_mom1} and Eq~\ref{eqn:perp_mom2}, that
\begin{equation}
    \frac{\partial \Mtot_s}{\partial t} + \nabla \cdot (\bm{V}_s \Mtot_s) + (\nabla \cdot \bm{P}_s)_\parallel \bvec - q_s n_s \bm{B} \times \vperpzeros = \bm{\mathcal{R}}_{\parallel s} + q_s n_s (\bm{E}_\parallel + \bm{V}_s \times \bm{B}).
\end{equation}
Finally, given that $q_s n_s \bm{B} \times \vperpzeros = q_s n_s \bm{E}_\perp - (\nabla \cdot \bm{P}_s)_\perp + \bm{\mathcal{R}}_{\perp s}$, we have
\begin{align}\label{eqn:momentum_density} 
     \frac{\partial \Mtot_s}{\partial t} + \nabla \cdot ( \bm{V}_s \Mtot_s + \bm{P}_s)  &= q_s n_s (\bm{E} + \bm{V}_s \times \bm{B}) + \bm{\mathcal{R}}_s ,
\end{align}
and, upon summing over all species $s$, we obtain the analogue of the non drift-reduced result, Eq~\ref{eqn:total_momentum_density}, that is
\begin{align}\label{eqn:LO_momentum}
    \frac{\partial \Mtot}{\partial t} + \nabla \cdot \sum_s ( \bm{V}_s \Mtot_s + \bm{P}_s)  &= \bm{J} \times \bm{B} + \sum_s \bm{\mathcal{R}}_s ,
\end{align}
showing that the leading-order contribution to the total fluid momentum is conserved.


 \section{Conclusion}\label{sec:Conclusion}

 The drift-reduced fluid model is a widely-used tool for simulating plasma turbulence in the collisional regime. Here, we derive a conservative drift-reduced system that holds in arbitrary magnetic geometry without enforcing the electrostatic limit, a crucial property for interpreting long-time dynamics and for constructing robust numerical solvers. The central step is a non-perturbative, analytic inversion of the defining relation for the polarisation velocity as a function of $\partial_t \bm{E}$, which yields a consistent transport equation for the leading-order perpendicular momentum and a closed set of conservative fluid equations.  The derivation presented here is independent of the closure and thus applies, for example, to Braginskii’s two-fluid closure as well as to multispecies closures \citep{Braginskii_1965, Zhdanov_2002, poulsen_ramussen_multispecies, Raghunathan_2022}. Unlike in variational approaches based on expanding the guiding-center Lagrangian~\citep{Brizard_variational_fluid_2005, Jorge_Ricci_Loureiro_2017, Mencke_Ricci_2025}, we performed the drift-reduction directly at the fluid level, starting from the equations of motion and requiring that conservation laws should be satisfied exactly. The implications of exact conservation properties on turbulence in drift-reduced fluid plasmas will be assessed in future work.

\section*{Acknowledgements}
The authors are grateful to D. Mancini, J. Mencke, L. Stenger and G. Van Parys for valuable discussions concerning drift-reduced models and guiding-center Lagrangian approaches, and thank the anonymous referees for their help in improving this manuscript. This work has been carried out within the framework of the EUROfusion Consortium, via the Euratom Research and Training Programme (Grant Agreement No 101052200 — EUROfusion) and funded by the Swiss State Secretariat for Education, Research and Innovation (SERI). Views and opinions expressed are however those of the author(s) only and do not necessarily reflect those of the European Union, the European Commission, or SERI. Neither the European Union nor the European Commission nor SERI can be held responsible for them. This work was supported
in part by the Swiss National Science Foundation.
inte\section*{Declaration of interest}
The authors report no conflict of interest.
\appendix

\section{Proof of $\vpols$ exact inversion}\label{sec:solution}

We prove that the expression of $\vpols$ in Eq~\ref{eqn:Gpol_solution} solves Eq~\ref{eqn:Gpol_start}. Since the computation is identical for all species, we omit the species index $s$ to avoid clutter. Substituting Eq~\ref{eqn:Gpol_solution} into Eq~\ref{eqn:Gpol_start}, we have
\begin{align}
    (\Delta - 1) \bm{U} = \frac{1}{\cycl}(\bvec \times \nabla \vzero) \bm{U} - \frac{1}{\cycl}\bvec \times (\bm{U} \cdot \nabla) \vzero - \frac{1}{\cycl^2}\bvec \times[((\bvec \times \nabla \vzero)\bm{U})\cdot \nabla \vzero] . 
\end{align}
Expressing the above in component form
\begin{align}
    (\Delta - 1) U^k = \frac{1}{\cycl}\epsilon^{k m n} b_m \nabla_n v_s U^s - \frac{1}{\cycl}\epsilon^{k m n} b_m \nabla_s v_n U^s - \frac{1}{\cycl^2}\epsilon^{k m n} b_m \epsilon^{pqr} b_q \nabla_p v_n \nabla_r v_s U^s .
\end{align}
Progress can be made by factorising one of the Levi-Civita terms and introducing again the tensor $\epsilon_{\perp}^{ij}$. We have  
\begin{align}\label{eqn:proof_partial}
    (\Delta - 1) U^k = -\frac{1}{\cycl}\epsilon_\perp^{kn} \left[\nabla_n v_s  -  \nabla_s v_n +  \frac{1}{\cycl} \epsilon_\perp^{pr}\nabla_p v_n \nabla_r v_s\right] U^s \equiv -\frac{1}{\cycl}\epsilon_\perp^{kn} L_{ns} U^s .
\end{align}
We note that the tensor
\begin{align}\label{eqn:L_tensor}
    L_{ns} = \nabla_n v_s  -  \nabla_s v_n +  \frac{1}{\cycl} \epsilon_\perp^{pr} \nabla_p v_n \nabla_r v_s,
\end{align}
is antisymmetric, $L_{ns} = - L_{sn}$. Indeed
\begin{align}
    L_{sn} &= \nabla_n v_s  -  \nabla_s v_n +  \frac{1}{\cycl} \epsilon_\perp^{pr} \nabla_p v_s \nabla_r v_n = \nabla_s v_n  -  \nabla_n v_s -  \frac{1}{\cycl} \epsilon_\perp^{rp} \nabla_r v_n \nabla_p v_s = -L_{ns} .
\end{align}
On $W_\perp$, any antisymmetric tensor takes the form $L_{ns} = \eta \epsilon_{\perp ns}$, for some scalar $\eta$. To compute $\eta$, we contract Eq~\ref{eqn:L_tensor} with $\epsilon_\perp^{ns}$. Recalling that $\epsilon_{\perp ns}\epsilon_\perp^{ns} =2$, we obtain
\begin{align}
    \eta &= \frac{1}{2} L_{ns} \epsilon_\perp^{ns} = \frac{1}{2} (\nabla_n v_s  -  \nabla_s v_n) \epsilon_\perp^{ns} +  \frac{1}{\cycl} \epsilon_\perp^{pr} \epsilon_\perp^{ns} \nabla_p v_n  \nabla_r v_s  = \epsilon_{\perp}^{ns}\nabla_n v_s +  \frac{1}{\cycl} \epsilon_\perp^{pr} \epsilon_\perp^{ns} \nabla_p v_n  \nabla_r v_s .
\end{align}
Expressing this result in terms of the trace of $\bvec \times \nabla \vzero$ and the determinant of $(\nabla \vzero)_\perp$ as was done in Eq~\ref{eqn:trace_levi} and~\ref{eqn:determinant_levi}, we have   
\begin{align}
    \eta = (\bvec \times \nabla) \cdot \vzero + \frac{1}{\cycl}\text{det}(\nabla \vzero)_\perp = \cycl (\Delta -1) .
\end{align}
Substituting the expression $L_{ns} = \eta \epsilon_{\perp ns}$, with $\eta = \cycl(\Delta -1)$ into Eq~\ref{eqn:proof_partial}, we find
\begin{align}
    U^k = - \epsilon_{\perp}^{kn} \epsilon_{\perp ns} U^s = (\Pi_\perp)^k_{\,\, s} U^s = U^k,
\end{align}
completing the proof.

\bibliographystyle{jpp}

\bibliography{jpp-instructions}

\end{document}